**An efficient estimation of time-varying parameters of dynamic models by combining offline batch optimization and online data assimilation**

Yohei Sawada[1],

[1] Institute of Engineering Innovation, the University of Tokyo, Tokyo, Japan

Corresponding author: Y. Sawada, Institute of Engineering Innovation, the University of Tokyo, Tokyo, Japan, 2-11-6, Yayoi, Bunkyo-ku, Tokyo, Japan, yohei.sawada@sogo.t.u-tokyo.ac.jp

**Abstract**

It is crucially important to estimate unknown parameters in earth system models by integrating observation and numerical simulation. For many applications in earth system sciences, an optimization method which allows parameters to temporally change is required. In the present paper, an efficient and practical method to estimate the time-varying parameters of relatively low dimensional models is presented. In the newly proposed method, called Hybrid Offline Online Parameter Estimation with Particle Filtering (HOOPE-PF), an inflation method to maintain the spread of ensemble members in a sampling-importance-resampling particle filter is improved using a non-parametric posterior probabilistic distribution of time-invariant parameters obtained by comparing simulated and observed climatology. The HOOPE-PF outperforms the original sampling-importance-resampling particle filter in synthetic experiments with toy models and a real-data experiment with a conceptual hydrological model especially when the ensemble size is small. The advantage of HOOPE-PF is that its performance is not greatly affected by the size of perturbation to be added to ensemble members to maintain their spread while it is critically important to get the optimal performance in the original particle filter. Since HOOPE-PF is the extension of the existing particle filter which has been extensively applied to many earth system models such as land, ecosystem, hydrology, and paleoclimate reconstruction, the HOOPE-PF can be applied to improve the simulation of these earth system models by considering time-varying model parameters.

**Plain Language Summary**

Computer simulation is widely used to understand earth systems such as atmosphere, ocean, land, ecosystem, and society. Many computer simulation models in earth system sciences inevitably have unknown coefficients of equations, which are called model parameters. It is crucially important for accurate simulation of earth systems to tune these model parameters by comparing the results of computer simulation with observation. Although many previous works assumed that model parameters do not change over time, it is necessary to allow them to change over time in some applications, which makes it more difficult to estimate model parameters. In this study, a new method to estimate time-varying model parameters is proposed. The key idea of this paper is to combine two existing methods which integrate observations into computer simulation models. It is found that the proposed method works better than the existing method in three case studies. The proposed method can contribute to improving many simulations of earth systems by efficiently tuning time-varying model parameters.

**Key Points**

An efficient and practical method to estimate time-varying model parameters is proposed.

Particle filtering can be stabilized by combining offline batch optimization and online data assimilation.

The new method is successfully applied to the Lorenz 63 model and the conceptual hydrological model.

## 1. Introduction

In Earth system sciences, numerical models are crucially important tools to understand, monitor, and predict earth systems. Many numerical models in earth system sciences have many unknown parameters which cannot be directly determined by current theory and cannot be directly measured. This uncertainty in model parameters substantially affects the skill of numerical models to simulate real phenomena. It is a grand challenge to infer model parameters by integrating observations and models. Although model parameters are often assumed to be time-invariant, there are time-varying parameters in practical applications due to incomplete parameterizations and unconsidered dynamics that control parameters (e.g., Reichert et al. 2021). It is beneficial to develop an efficient and practical parameter estimation method which allows parameters to temporally change.

To estimate time-varying parameters, online parameter estimation by sequential data assimilation has been recognized as a useful method. Since ensemble data assimilation such as Ensemble Kalman filter (EnKF) and Particle Filter (PF) can sequentially adjust model parameters using real-time observations, it can be easily applied to the estimation of time-varying parameters. Ruiz et al. (2013a) applied the Local Ensemble Transform Kalman Filter (LETKF; Hunt et al. 2007) to jointly estimate state variables and parameters in a simple low-resolution General Circulation Model (GCM). They successfully estimated the time-varying parameters by sequentially adjusting them in an idealized experiment. Pathiraja et al. (2018) successfully detected a land use change in a river basin by applying the locally linear dual EnKF proposed by Pathiraja et al. (2016) to estimate a temporal change in parameters of hydrological models. Sawada and Hanazaki (2020) applied PF to a socio-hydrological model proposed by Di Baldassarre et al. (2013), in which flood-human interactions are simulated. They successfully estimated time-varying parameters driven by unknown social dynamics in an idealized experiment.

Whenever parameters are adjusted by ensemble data assimilation, the estimated ensemble variance of updated parameters is always smaller than that of the background. If the ensemble variance is too small compared to observation error variances, the information of observation cannot impact the state and parameter estimation. To avoid this filter degeneracy (Anderson 2007), it is crucially important to maintain the appropriate ensemble variance of updated parameters. Many parameter perturbation methods or inflation methods have been proposed to maintain the ensemble variance of estimated parameters. Moradkhani et al. (2005a) simply added the Gaussian noise, whose variance is proportional to the ensemble variance of the background, to parameters of each ensemble member after adjusting them by the sampling-importance-resampling PF. Moradkhani et al. (2005b) used the kernel smoother with location shrinkage (Liu and West 2001) to perturb updated parameters and maintain the ensemble variance of parameters in their dual EnKF. Similar strategies to Moradkhani et al. (2005a, 2005b) have been used in many previous works (e.g., Yan and Moradkhani 2016; Pathiraja et al. 2016; Ait-El-Fquih and Hoteit 2017; Pathiraja et al. 2018; Sawada and Hanazaki 2020). It has been shown effective to evaluate the perturbed parameters by running a numerical model again and iteratively sampling the new perturbed parameters (e.g., Moradkhani et al. 2012; Vrugt et al. 2013; Abbaszadeh et al. 2018), although the computational cost is increased by this strategy. Ruiz et al. (2013) discussed how to objectively determine the appropriate variance of updated parameters using the ensemble variance of background and updated state variables. Kotsuki et al. (2018) used the relaxation to prior spread method (Whitaker and Hamil 2012) to constrain the ensemble variance of parameters to initial prescribed variance in their LETKF framework.

As an alternative approach of the parameter estimation, offline batch optimization, in which model parameters are optimized by minimizing a cost function based on the difference between simulation and observation in a long-term (e.g., longer than a year in hydrology) period, has been used. The advantage of the offline batch optimization is that non-parametric Bayesian inference can easily be applied in this framework, so that the reliable uncertainty estimation of model parameters can be obtained. Since it is necessary to iteratively perform the long-term integration of a numerical model with different parameters, the direct application of the offline batch optimization to models in earth system sciences is computationally expensive (e.g., Laloy and Vrugt 2012). However, the computational cost of the offline batch optimization can be dramatically reduced by replacing the responses of numerical models to parameters with machine learning models called surrogate models. For example, Sawada (2020) used the Gaussian process regression (Rasmussen and Williams 2006) to mimic the relationship between parameters of a land surface model and simulated microwave brightness temperature which is sensitive to soil moisture and vegetation water content. By evaluating parameters based on the computationally cheap statistical surrogate model, the Markov Chain Monte Carlo (MCMC) sampler (Hastings 1970) could be efficiently applied to optimize soil and vegetation parameters of the land surface model. Zhang et al. (2020) proposed a method to reduce the approximation error of a surrogate model by adding the samples based on the evaluation of the surrogate. Their method was successfully applied to the parameter inference of a groundwater model. Parente et al. (2019) applied the active subspace method (Constantine et al. 2014) to efficiently construct a surrogate model and calibrate parameters of a hydrological model. Dunbar et al. (2021) successfully performed the surrogate model based MCMC and realized the uncertainty quantification of convective parameters in an idealized GCM with only $O(10^2)$ model runs, which is relatively cheap considering that the direct application of MCMC requires $O(10^6)$ model runs.

Although several studies proposed combining data assimilation and offline batch optimization (e.g., Cleary et al. 2021; Lunderman et al. 2021; Tomizawa and Sawada 2021), there is no contribution to the estimation of time-varying parameters by combining them. The aim of this study is to develop an efficient and practical model parameter optimization method which can allow parameters to temporally vary by combining online PF, which has been widely used in the earth system sciences (e.g., Qin et al. 2009; Mairesse et al. 2013; Sawada et al. 2015; Abolafia-Rosenzweig et al. 2019), with offline batch optimization. The proposed method substantially reduces the sensitivity of the PF's performance to a hyperparameter of a parameter perturbation method. Therefore, the proposed method contributes to solving the issues in the maintenance of ensemble variance discussed above. This feature will be demonstrated in idealized experiments with the Lorenz 63 model (Lorenz 1963) and a real-data experiment with the HYMOD conceptual hydrological model (Moore 2007).

## 2. Method
### 2.1. Online data assimilation: Particle Filtering
In this study, the Sampling-Importance-Resampling Particle Filter (SIRPF) algorithm was used as a data assimilation method. The implementation of Moradkhani et al. (2005a) was mostly used in this study. Although Moradkhani et al. (2005a) used the effective sample size to determine if a perturbation of ensembles is implemented, this method was not adopted in this study. A discrete state-space dynamics system is defined as follows:

$$\boldsymbol{x}(t) = f\big(\boldsymbol{x}(t-1), \boldsymbol{\theta}(t-1), \boldsymbol{u}(t-1)\big) + \boldsymbol{q}(t-1) \quad (1)$$

where $\boldsymbol{x}(t)$ is the state variables, $\boldsymbol{\theta}(t)$ is the model parameters, $\boldsymbol{u}(t)$ is the external forcing, and $\boldsymbol{q}(t)$ is the noise process which represents the model error at time *t*. *f(.)* is the dynamic model. Note that the model parameters are the function of *t*, so that it is assumed that time-varying parameters exist. Since observable variables are not fully identical to the state variables of the dynamic model in general cases, data assimilation needs to formulate an observation process:

$$y^f(t) = h(x(t)) + r(t) \quad (2)$$

where $y^f(t)$ is the simulated observation at time $t$, $r(t)$ is the noise process which represents observation error at time $t$. $h(.)$ is the observation operator. The purpose of PF (and any other sequential data assimilation methods) is to obtain the posterior distribution of the state variables and parameters based on the Bayesian update:

$$p(x(t), \theta(t)|y^o(1:t)) \propto p(y^o(t)|x(t), \theta(t)) p(x(t), \theta(t)|y^o(1:t-1)) \quad (3)$$

where $p(x(t), \theta(t)|y^o(1:t))$ is the posterior probability of the state variables $x(t)$ and parameters $\theta(t)$ given all observation up to time $t$, $y^o(1:t)$. The prior knowledge, $p(x(t), \theta(t)|y^o(1:t-1))$, is updated using the likelihood, $p(y^o(t)|x(t), \theta(t))$ which includes the information of the new observation at time $t$. PF directly approximates equation (3) by Monte Carlo simulation.

The implementation of SIRPF in this study is as follows:

*1. Forecasting model state variables and parameters from time t-1 to t using ensemble simulation.*

$$x_i^f(t) = f(x_i^a(t-1), \theta_i^a(t-1), u(t-1)) \quad (4)$$
$$\theta_i^f(t) = \theta_i^a(t-1) \quad (5)$$

where superscripts *f* and *a* show forecasted and updated variables, respectively. Subscript *i* shows the ensemble number. Equation (5) implies that the model parameters do not change in the forecast step.

*2. Calculating simulated observations for all ensembles:*

$$y_i^f(t) = h\left(x_i^f(t)\right) \quad (6)$$

*3. Calculating weights for all ensembles:*

$$w_i = \frac{p(y^o(t)|x_i^f(t), \theta_i^f(t))}{\sum_{k=1}^N p(y^o(t)|x_k^f(t), \theta_k^f(t))} \quad (7)$$

where $w_i$ is the normalized weight for the *i*th ensemble member, and $N$ is the ensemble size. $p(y^o(t)|x_i^f(t), \theta_i^f(t))$ is the likelihood function. In this study, it is assumed that the observation error follows the Gaussian distribution, so that $p(y^o(t)|x_i^f(t), \theta_i^f(t))$ can be formulated as follows:

$$p(y^o(t)|x_i^f(t), \theta_i^f(t)) = \frac{1}{\sqrt{\det(2\pi R)}} \exp\left(-\frac{1}{2}(y^o(t) - y_i^f(t))^T R^{-1}(y^o(t) - y_i^f(t))\right)$$

(8)

where **R** is the covariance matrix of the observation error.

*4. Resampling ensemble members based on normalized weights.* The normalized weight $w_i$ is recognized as the probability that the ensemble i is selected after resampling.

Multinomial resampling (Gordon et al. 1993) was used as a resampling method. The resampled state variables and parameters are defined as $x_i^{resamp}$ and $\theta_i^{resamp}$, respectively.

*5. Adding a perturbation to state variables and parameters to maintain appropriate ensemble variance:*

$x_i^a = x_i^{resamp} + \varepsilon_{i,state}$ (9)

$\varepsilon_{i,state} \sim N(0, S_{state} \times Var^x)$ (10)

$\theta_i^a = \theta_i^{resamp} + \varepsilon_{i,para}$ (11)

$\varepsilon_{i,para} \sim N(0, S_{para} \times Var^\theta)$ (12)

where N(.) is the Gaussian distribution, $Var^x$ and $Var^\theta$ are the ensemble variances of the state variables $x_i^f(t)$ and parameters $\theta_i^f(t)$ in the background field, respectively.

$S_{state}$ and $S_{para}$ are the "inflation" hyperparameters. The performance of the original PF is sensitive to them. Several approaches to determine them objectively and adaptively have been proposed (e.g., Leisenring and Moradkhani 2012; Sawada and Hanazaki 2020). In this study, these adaptive approaches were not used and these hyperparameters were fixed. The experiments of this study will show that the performance of a filter is less sensitive to these hyperparameters in the proposed method than the original SIRPF. Note that these adaptive approaches cannot eliminate hyperparameters in the filter since another new hyperparameters appear in the adaptive method. Our proposed method may also be able to reduce the importance of these hyperparameters in the adaptive methods although it is not the scope of the present paper.

**2.2. Offline batch optimization**

The aim of the offline batch optimization used in this paper is to get a probabilistic distribution to check the quality of the perturbed parameters. Note that the offline batch optimization of this study is not used to get the best set of parameters (see also section 2.3). In the offline batch optimization of this paper, the parameters are assumed to be time-invariant. A discrete state-space dynamics system is re-written as:

$x(t) = f\left(x(t-1), \widehat{\theta}, u(t-1)\right) + q(t-1)$ (13)

$y^f(t) = h(x(t)) + r(t)$ (14)

where $\widehat{\theta}$ is the time-invariant model parameter. In this study, some climatological indices were calculated based on the long-term averages of the timeseries of the simulated observation $y^f(t)$ and the real observation $y^o(t)$. Since these climatological indices are designed to be independent to the initial condition, the problem of the offline batch optimization can be simplified as:

$$\gamma^f = g(\widehat{\boldsymbol{\theta}}) \quad (15)$$

where $\gamma^f$ is the simulated climatological index. $g(.)$ is the forward map. The purpose of the offline batch optimization in this study is to obtain the posterior probabilistic distribution of $\widehat{\boldsymbol{\theta}}$, which is defined as $Q(\widehat{\boldsymbol{\theta}})$, based on the simulated climatological index $\gamma^f$ and the observed climatological index $\gamma^o$.

It is generally difficult to perform the Bayesian inference using equation (15) since $g(.)$ is computationally expensive. In this study, the surrogate model of g(.) was developed by the Gaussian process regression (Rasmussen and Williams 2006). First, 500 ensemble members of model parameters were generated by the pseudo-Monte Carlo sampling which enables to uniformly draw samples from the multi-dimensional parameter space. Second, the long-term integration of the dynamic models (i.e. the Lorenz 63 and HYMOD models) with the 500 parameter ensemble members was performed in parallel. Third, the statistical surrogate model was constructed from the data of the 500 $\widehat{\boldsymbol{\theta}}$-$\gamma^f$ combinations by the Gaussian process regression. The surrogate model replaces equation (15) with:

$$\gamma^f = g^{(s)}(\widehat{\boldsymbol{\theta}}) \quad (16)$$

where $g^{(s)}(.)$ is the computationally cheap surrogate model of $g(.)$. While the long-term integration of a full dynamic model with many input data is necessary to obtain its climatology and evaluate $g(\widehat{\boldsymbol{\theta}})$, the trained surrogate model $g^{(s)}(\widehat{\boldsymbol{\theta}})$ requires only model parameters and directly estimates the climatology without performing the long-term integration. Therefore, the evaluation of equation (16) is much cheaper and less complex than that of equation (15) if the surrogate model can accurately mimic the responses of a dynamic model to parameters. The amount of the training data (i.e., ensemble size) affects the skill of the Gaussian process regression to mimic the responses of a dynamic model to its parameters. The skill of the Gaussian process regression to simulate $\gamma^f$ was evaluated by the independent test set of 1,000 simulation of dynamic models. In all applications of this paper (see also section 3), the correlation coefficient between predicted and true $\gamma^f$ is more than 0.95, so that it is confirmed that the 500-ensemble size is enough in this study.

The Metropolis-Hastings algorithm (Hastings 1970) was used as a MCMC sampler to draw samples from the probabilistic distribution $Q(\widehat{\boldsymbol{\theta}})$. The implementation of the MCMC sampler is the following:

*1. For each iteration i, generate a candidate parameter vector $\widehat{\boldsymbol{\theta}}_c$.* This candidate is sampled from the proposal distribution $q(\widehat{\boldsymbol{\theta}}_c|\widehat{\boldsymbol{\theta}}_i)$.

*2. Calculating an acceptance probability of $\widehat{\boldsymbol{\theta}}_c$, $\alpha(\widehat{\boldsymbol{\theta}}_i, \widehat{\boldsymbol{\theta}}_c)$:*

$$\alpha(\widehat{\boldsymbol{\theta}}_i, \widehat{\boldsymbol{\theta}}_c) = \exp\left(\Phi^{(s)}(\widehat{\boldsymbol{\theta}}_i) - \Phi^{(s)}(\widehat{\boldsymbol{\theta}}_c)\right) \quad (17)$$

$$\Phi^{(s)}(\widehat{\boldsymbol{\theta}}) = \frac{1}{2}\left\|\boldsymbol{\gamma}^o - g^{(s)}(\widehat{\boldsymbol{\theta}})\right\|^2_{R_{gp}(\widehat{\boldsymbol{\theta}}) + R_o} \quad (18)$$

where $R_{gp}(\widehat{\boldsymbol{\theta}})$ and $R_o$ are the error variances of $g^{(s)}(\widehat{\boldsymbol{\theta}})$ and observation, respectively (see also Cleary et al. 2020). The acceptance probability is calculated based on the square difference between the simulated and observed climatological indices normalized by the total error variance of the Gaussian process and the observation measurement.

*3. Determine if $\widehat{\boldsymbol{\theta}}_c$ is accepted as a new parameter or not.* A random number, $b$, is generated from the uniform distribution of [0,1]. Then,

If $b \leq \alpha(\widehat{\boldsymbol{\theta}}_i, \widehat{\boldsymbol{\theta}}_c)$, accept the candidate parameter and $\widehat{\boldsymbol{\theta}}_{i+1} = \widehat{\boldsymbol{\theta}}_c$

If $b > \alpha(\widehat{\boldsymbol{\theta}}_i, \widehat{\boldsymbol{\theta}}_c)$, reject the candidate parameter and $\widehat{\boldsymbol{\theta}}_{i+1} = \widehat{\boldsymbol{\theta}}_i$

These three steps were iterated 500,000 times in this study and the first 100,000 iterations were discarded as the spin-up period. From the remaining 400,000 samples, the probabilistic distribution of parameters $Q(\widehat{\boldsymbol{\theta}})$ was obtained. In this study, $q(\widehat{\boldsymbol{\theta}}_c|\widehat{\boldsymbol{\theta}}_i)$ is assumed to be Gaussian with zero mean. Since the Gaussian process can provide both mean and variance of their estimation in each point, it is straightforward to obtain $R_{gp}(\widehat{\boldsymbol{\theta}})$ whenever $g^{(s)}(\widehat{\boldsymbol{\theta}})$ is evaluated. To calculate $R_o$, the subset of the continuous timeseries of observations was randomly chosen and 1,000 observed climatological indices $\boldsymbol{\gamma}^o$ were generated. The variance of the observed climatological index was calculated as the variance of these 1,000 $\boldsymbol{\gamma}^o$. In addition, $\boldsymbol{\gamma}^o$ was randomly chosen from these subsets of the observed climatological indices in every 100 iterations of MCMC. This procedure was found to be necessary to obtain the posterior of the time-invariant $\widehat{\boldsymbol{\theta}}$ when parameters actually change over time and make observed variables non-stationary. If this process is neglected and $R_o$ is set to zero, the posterior distribution of $\widehat{\boldsymbol{\theta}}$ becomes underdispersive although the relatively large uncertainty of $\widehat{\boldsymbol{\theta}}$ needs to be obtained since the actual parameters temporally change.

*2.3. Hybrid Offline Online Parameter Estimation with Particle Filtering (HOOPE-PF)*
The schematic of the proposed method, Hybrid Offline Online Parameter Estimation with Particle Filtering (HOOPE-PF), is shown in Figure 1. The fundamental idea is that the posterior of PF, $p(x(t), \theta(t)|y^o(1:t))$ (see section 2.1), is constrained by the probabilistic distribution of parameters from the offline batch optimization, $Q(\hat{\theta})$ (see section 2.2) in HOOPE-PF. The parameter perturbation method was slightly changed to include the information of $Q(\hat{\theta})$ into the process of the parameter perturbation after resampling. The implementation of HOOPE-PF is the following. Note that the steps from 1 to 5a are the same as the original SIRPF described in the section 2.1.

*0. Performing the offline batch optimization to get the probability distribution of time-invariant parameters $Q(\hat{\theta})$.*

*1. Forecasting model state variables and parameters from time t-1 to t using ensemble simulation.*

$$x_i^f(t) = f(x_i^a(t-1), \theta_i^a(t-1), u(t-1)) \quad (4)$$

$$\theta_i^f(t) = \theta_i^a(t-1) \quad (5)$$

where superscripts *f* and *a* show forecasted and updated variables, respectively. Subscript *i* shows the ensemble number. Equation (5) implies that the model parameters do not change in the forecast step.

*2. Calculating simulated observations for all ensembles:*

$$y_i^f(t) = h\left(x_i^f(t)\right) \quad (6)$$

*3. Calculating weights for all ensembles:*

$$w_i = \frac{p(y^o(t)|x_i^f(t), \theta_i^f(t))}{\sum_{k=1}^{N} p(y^o(t)|x_k^f(t), \theta_k^f(t))} \quad (7)$$

where $w_i$ is the normalized weight for the *i*th ensemble member, and *N* is the ensemble size. $p(y^o(t)|x_i^f(t), \theta_i^f(t))$ is the likelihood function. In this study, it is assumed that the observation error follows the Gaussian distribution, so that $p(y^o(t)|x_i^f(t), \theta_i^f(t))$ can be formulated as follows:

$$p(y^o(t)|x_i^f(t), \theta_i^f(t)) = \frac{1}{\sqrt{\det(2\pi R)}} \exp\left(-\frac{1}{2}(y^o(t) - y_i^f(t))^T R^{-1}(y^o(t) - y_i^f(t))\right)$$

(8)

where **R** is the covariance matrix of the observation error.

*4. Resampling ensemble members based on normalized weights.* The normalized weight $w_i$ is recognized as the probability that the ensemble i is selected after resampling. Multinomial resampling (Gordon et al. 1993) was used as the resampling method. The resampled state variables and parameters are defined as $x_i^{resamp}$ and $\theta_i^{resamp}$, respectively.

*5a. Adding a perturbation to state variables and parameters to maintain appropriate ensemble variances:*

$$x_i^a = x_i^{resamp} + \varepsilon_{i,state}$$

$$\varepsilon_{i,state} \sim N(0, S_{state} \times Var^x)$$

$$\theta_i^a = \theta_i^{resamp} + \varepsilon_{i,para}$$

$$\varepsilon_{i,para} \sim N(0, S_{para} \times Var^\theta)$$

where N(.) is the Gaussian distribution, $Var^x$ and $Var^\theta$ are the ensemble variances of the state variables $x_i^f(t)$ and parameters $\theta_i^f(t)$ in the background field, respectively.

*5b. Calculating an acceptance probability of $\theta_i^a$, $\alpha(\theta_i^{resamp}, \theta_i^a)$ by evaluating how $\theta_i^a$ corresponds to the results of the offline batch optimization, $Q(\hat{\theta})$:*

$$\alpha(\theta_i^{resamp}, \theta_i^a) = \frac{Q(\theta_i^a)}{Q(\theta_i^{resamp})} \quad (19)$$

Here Q(.) is used as a function to evaluate how a parameter vector fits the posterior distribution of $\hat{\theta}$ obtained by the offline batch optimization. Equation (19) compares the fitness of the perturbed parameter $\theta_i^a$ to the posterior distribution of $\hat{\theta}$ with that of the original parameter $\theta_i^{resamp}$.

*5c. Determine if $\theta_i^a$ is accepted as a new parameter or not.* A random number, *b*, is generated from the uniform distribution of [0,1]. Then,

If $b \leq \alpha(\theta_i^{resamp}, \theta_i^a)$, accept the updated parameter $\theta_i^a$

If $b > \alpha(\theta_i^{resamp}, \theta_i^a)$, reject the current updated parameter and go back to the step 5 only for the parameters.

The new steps 5b and 5c can prevent from sampling parameters which cannot support the results of the offline batch optimization. If the fitness of the perturbed parameter to the results of the offline batch optimization, $Q(\theta_i^a)$, is substantially lower than that of the original resampled parameter $Q(\theta_i^{resamp})$, this perturbed parameter is unlikely to be chosen. In the original SIRPF, large $S_{para}$ moves particles far away from the resampled ones in any directions, which makes the filter unstable. Although large $S_{para}$ of HOOPE-PF also moves particles far away from the resampled ones, it tends to move particles in the direction which increases $Q(.)$ since the perturbation which reduces $Q(.)$ tends to be rejected in the step 5c. The numerical experiments in the following sections will reveal that this new perturbation method substantially stabilizes PF. Note that this new perturbation method can be used for any kinds of $Q(.)$, and $Q(.)$ is not necessarily obtained by the methods described in the section 2.2. It is possible to obtain $Q(.)$ by computationally cheaper methods such as assuming the Gaussian distribution. This study suggested that $Q(.)$ is obtained by the fully nonparametric MCMC method to maximize the potential of PF to deal with the non-Gaussian probabilistic distribution.

## 3. Experiment design

### 3.1. Case study I: The Lorenz 63 model with an abruptly changed parameter

HOOPE-PF was tested with the Lorenz 63 model (Lorenz 1963):

$$\frac{dx}{dt} = -10(x - y) \quad (20)$$

$$\frac{dy}{dt} = -xz + \rho(t)x - y \quad (21)$$

$$\frac{dz}{dt} = xy - bz \quad (22)$$

$\rho(t)$ and $b$ are unknown parameters to be estimated. The true b is assumed to 8/3. Although the parameter in equation (21) is time-invariant in the original Lorenz 63 model, here it is assumed that $\rho(t)$ was a time-varying parameter:

$$\rho(t) = \begin{cases} 28 \ (0 \leq t < 8000, 16000 \leq t < 24000) \\ 24 \ (8000 \leq t < 16000, 24000 \leq t < 32000) \end{cases} \quad (23)$$

In this case study, $\rho(t)$ abruptly changes every 8,000 timesteps. The size of the timestep was set to 0.01 and equations (20-22) were numerically solved by the 4th-order Runge-Kutta scheme.

The two state variables, *y* and *z*, are assumed to be observed every 20 timesteps and the observation error was set to 1.0. The climatological index $\boldsymbol{\gamma} = [\overline{y^2}, \overline{z^2}]$ was calculated by averaging simulated and observed variables during 4,000 timesteps. The 500 ensemble runs with different (time-invariant) parameters were performed for 4,000 timesteps to construct the surrogate model ($g^{(s)}(\widehat{\boldsymbol{\theta}})$). In the MCMC sampler, the observations up to *t*=32000 were used to generate the subsets of the observation timeseries (see section 2.2). The period of each subset corresponded to 4000 timesteps of the model. The initial conditions of state variables were generated by adding the Gaussian white noise whose mean and standard deviation are 0 and 1, respectively to the true values. The initial conditions of $\rho(t)$ were drawn from the uniform distribution from 10 to 40. The initial conditions of *b* were drawn from the uniform distribution from 0 to 15. The performance of HOOPE-PF was compared with that of the original PF. The other settings of the hyperparameters and the ensemble size can be found in Table 1. The ensemble median was used as a representative value of the estimated probabilistic distribution to evaluate the filters because the median is often consistent to the mode of the probabilistic distribution in this experiment.

### 3.2. Case study II: The Lorenz 63 model with non-periodic forcing

In this second case study, the Lorenz 63 model described in equations (20-22) was also used. However, the time-varying parameter was defined as:

$\rho(t) = \rho_o + \psi(t)$ (24)

$\psi(t) = A(\frac{1}{3}\sin(2\pi f t) + \frac{1}{3}\sin(\sqrt{3} f t) + \frac{1}{3}\sin(\sqrt{17} f t))$ (25)

where $\rho_o = 28$, $A = 5$, and $f = 1/20$. This version of the Lorenz 63 model has been intensively analyzed by Daron and Stainforth (2015). The settings of observation and a climatological index are the same as the case study I. The performance of HOOPE-PF was compared with that of the original PF. The other settings of the hyperparameters and the ensemble size can be found in Table 1.

### 3.3. Case study III: Real-data experiment with the conceptual hydrological model

While the previous two case studies are the synthetic experiments, here the parameters of the conceptual hydrological model, HYMOD (Moore 2007), were estimated by assimilating the real runoff observation. HYMOD has been widely used for studies on parameter optimization (e.g., Moradkhani et al. 2005a, Vrugt et al. 2013; Pathiraja et al. 2018). HYMOD consists of a nonlinear rainfall excess model connected to four linear routing reservoirs (three quick-flow reservoirs and a slow-flow reservoir). There are 5 state variables and 5 parameters in HYMOD. Table 2 shows the description and ranges of these 5 parameters.

The study area is the Leaf River basin, Mississippi, the United States. The Leaf River basin has been used to test data assimilation and parameter optimization methods by previous works (e.g., Moradkhani et al. 2005a; Vrugt et al. 2013; Sheikholeslami and Razavi 2020). The data of precipitation, potential evapotranspiration, and runoff were extracted from the Catchment Attributes and Meteorology for Large Sample studies dataset (CAMELS; Addor et al. 2017).

The climatological indices were set to the 5-year runoff ratio and the 5-year baseflow index. The runoff ratio is defined as the ratio of total runoff to total precipitation. The baseflow index is defined as the ratio of total baseflow to total runoff. Baseflow separation was performed by the one-parameter Lyne-Hollick digital filter (Lyne and Hollick 1979).

Following Gnann et al. (2019), the Lyne-Hollick digital filter was applied forward, backward, and forward again using a filter parameter of 0.925.

The 500 ensemble runs were performed for 5 years (from 1981-1985) to construct the surrogate model. This surrogate model directly estimates runoff ratio and baseflow index from the 5 parameters of HYMOD. In the MCMC sampler, observations from 1981-2000 were used to generate the subsets of the observation timeseries (see section 2.2) and the period of each subset was set to 5-year. Although the HOOPE-PF was performed from 1981 to 2014, the results from 1981-2000 were not used for evaluation. The performance of HOOPE-PF was evaluated from 2001 to 2014 so that the observation data used for the offline batch optimization were not used for the evaluation of HOOPE-PF.

When the state variables are updated, the logarithm of them was taken to prevent from obtaining negative values of the state variables. The error variance of observed runoff was set to 10% of the observed value following Pathiraja et al. (2018). In addition, the minimum observation variance was set to 0.1 since the filter degeneracy was induced by too small observation errors. The state variables and parameters were updated by assimilating the daily runoff observation every day. The performance of HOOPE-PF was compared with the original PF. The other settings of the hyperparameters and the ensemble size can be found in Table 1.

The initial conditions of the state variables for each ensemble were set to 0. The initial conditions of parameters for each ensemble were drawn from the uniform distribution whose upper and lower bounds were shown in Table 2.

The assimilated daily runoff was also used to evaluate the model simulation. The performance was evaluated by Kling-Gupta Efficiency (KGE, Gupta et al. 2009):

$$KGE = 1 - \sqrt{(r-1)^2 + (\frac{\sigma_{sim}}{\sigma_{obs}} - 1)^2 + (\frac{\mu_{sim}}{\mu_{obs}} - 1)^2} \quad (26)$$

where $r$ is a correlation coefficient between simulated and observed runoff, $\sigma_{sim}$ and $\sigma_{obs}$ are standard deviations of simulated and observed runoff, respectively. $\mu_{sim}$ and $\mu_{obs}$ are mean of simulated and observed runoff, respectively. KGE changes from $-\infty$ to 1 and the larger KGE means the better skill to reproduce observed runoff. KGE=1 indicates perfect correspondence between simulated and observed runoff. The ensemble median was used as a representative value of the estimated probabilistic distribution to evaluate the filters. Although the probabilistic distribution of the estimated runoff is highly skewed and does not follow the Gaussian distribution, the median is often consistent to the mode of the probabilistic distribution in this experiment.

## 4. Results
### 4.1. Case study I: The Lorenz 63 model with an abruptly changed parameter

Figure 2 shows the probabilistic distribution of parameters obtained by the offline batch optimization, $Q(\widehat{\boldsymbol{\theta}})$. Both $\rho$ and $b$ are distributed around the synthetic truth. Although most of observation information is not used by averaging its long-term timeseries in the offline batch optimization, the posterior distribution which reasonably covers the synthetic truth can be obtained.

In Figure 3, the performances of HOOPE-PF and PF are evaluated by Root-Mean-Square-Error (RMSE) of $\rho(t)$ as the function of the two inflation hyperparameters (equations 10 and 12). Since it is the synthetic experiment, the estimated parameter can be directly compared with the synthetic truth of the parameter. First, HOOPE-PF outperforms PF especially when the ensemble size is small (N=30). The minimum RMSE of HOOPE-PF with the ensemble sizes of 30, 100, and 250 are 0.93, 0.60, and 0.56, respectively. On the other hand, the minimum RMSE of PF with the ensemble sizes of 30, 100, and 250 are 1.21, 0.71, and 0.58, respectively. Second, in HOOPE-PF, the sensitivity of RMSE to the hyperparameter $S_{para}$ of $\rho(t)$ is much smaller than that in PF. Except for the case of the smallest $S_{para}$, HOOPE-PF stably provides the accurate estimation of $\rho(t)$ especially when the ensemble size is large (Figures 3e and 3f). On the other hand, the performance of PF is substantially sensitive to $S_{para}$ (Figures 3b and 3c). In the case of the large $S_{para}$, PF collapses. This superiority of HOOPE-PF to PF can also be found in RMSE of the state variables (results not shown).

Figure 4 shows the timeseries of $\rho(t)$ in the PF and HOOPE-PF with the selected settings to demonstrate how differently these two data assimilation methods work. The purpose of Figure 4 is to demonstrate how the estimated probabilistic distribution of $\rho(t)$ is affected by $S_{para}$ in the PF and HOOPE-PF. Note that the comparison of the performances between PF and HOOPE-PF is not the purpose of these figures. The superiority of HOOPE-PF over PF has already been shown in Figure 3. In the case of the small $S_{para}$, both PF and HOOPE-PF cannot accurately estimate the change in the parameter since the ensemble variance is too small to detect the abrupt regime change (Figures 4a and 4e). When $S_{para}$ increases to 0.5, both PF and HOOPE-PF maintain the appropriate ensemble variances and successfully estimate the time-varying parameter. However, in PF, the filter becomes unstable when $S_{para}$ further increases, and PF completely collapses in $S_{para}$ = 0.9 (Figures 4c and 4d). On the other hand, HOOPE-PF stably works in the cases of the higher $S_{para}$ (Figures 4g and 4h). This is because the perturbed parameters are constrained by the result of the offline batch optimization in HOOPE-PF. The constraint of the offline batch optimization prevents ensemble members from going to the unrealistic space of parameters in terms of model's climate even if each ensemble member is strongly perturbed by the large $S_{para}$.

### 4.2. Case study II: The Lorenz 63 model with non-periodic forcing

Figure 5 shows the probabilistic distribution of parameters obtained by the offline batch optimization, $Q(\hat{\theta})$ in the case study II. Since true $\rho(t)$ changes from 23 to 33 (equations (24-25)), the posterior distribution of $\rho(t)$ obtained by the offline batch optimization captures the range of the synthetic true parameter.

Figure 6 indicates that HOOPE-PF outperforms PF in this case study as found in the case study I. In HOOPE-PF, the change in $\rho(t)$ can be accurately estimated when $S_{para}$ is relatively large even if the ensemble size is small. The minimum RMSE in HOOPE-PF with the ensemble sizes of 30, 100, and 250 are 0.94, 0.70, and 0.65, respectively. On the other hand, the minimum RMSE in PF with the ensemble sizes of 30, 100, and 250 are 1.06, 0.74, and 0.68, respectively. The performance of PF is substantially sensitive to $S_{para}$ even in the cases of the large ensemble sizes. It implies that the intensive calibration of the inflation hyperparameter is necessary if one uses no objective estimation schemes of the hyperparameters. In HOOPE-PF, this intensive hyperparameter calibration is unnecessary. Note that some test runs to find proper hyperparameters are still necessary even in HOOPE-PF according to Figure 6.

Figure 7 shows the timeseries of $\rho(t)$ in PF and HOOPE-PF with the selected settings to demonstrate how differently these two data assimilation methods work. The purpose of Figure 7 is to demonstrate how the estimated probabilistic distribution of $\rho(t)$ is affected by $S_{para}$ in PF and HOOPE-PF. Note that the comparison of the performances between PF and HOOPE-PF is not the purpose of these figures. The superiority of HOOPE-PF over PF has already been shown in Figure 6. Figure 7 demonstrates that $S_{para}$ plays the important role in controlling the ensemble variance and the skill to stably estimate the time-varying parameter in the original PF. On the other hand, the offline batch optimization largely determines the ensemble variance in HOOPE-PF and the sensitivity of the performance of the parameter estimation to $S_{para}$ in HOOPE-PF is much smaller than that in PF.

**4.3. Case study III: Real-data experiment with the conceptual hydrological model**
Figure 8 shows the histograms of the HYMOD's 5 parameters (see Table 2) estimated by the offline batch optimization. All parameters are shown as normalized values by the prescribed maximum and minimum values (see Table2). There is no known synthetic truth in this case study since the real runoff observation was used in this experiment. The 5-year runoff ratio and baseflow index effectively constrain $b_{exp}$ and $k_s$. Although the relatively large variances remain in the other three parameters, all 5 parameters sampled by MCMC are mutually correlated with each other (not shown). Therefore, the offline batch optimization substantially reduces the total volume of the parameter space which should be searched by the particle filtering.

In Figure 9, the performances of HOOPE-PF and PF are evaluated by the Kling-Gupta Efficiency (KGE, Gupta et al. 2009) as the function of the two inflation hyperparameters. Figure 9 clearly reveals that HOOPE-PF outperforms PF in most cases. The maximum KGE in HOOPE-PF with the ensemble sizes of 30, 100, and 250 are 0.79, 0.80, and 0.84, respectively. On the other hand, the maximum KGE in PF with the ensemble sizes of 30, 100, and 250 are 0.47, 0.73, and 0.79, respectively. In the supporting information, the timeseries of runoff and $k_q$ (see Table 2) provided by the optimal hyperparameters which maximize KGE are shown (Figures S1-S3). The effects of the two hyperparameters on the stability and accuracy of the simulations in this case study are different from those in the previous case studies. The performance of PF is not degraded when $S_{para}$ is set to the large values. This is mainly because the model parameters are bounded in the reasonable prescribed ranges shown in Table 2, which prevents the filter from collapsing when the ensemble variance of parameters is strongly inflated. In this case study, ensemble members are already constrained by the prescribed ranges shown in Table 2, which stabilizes the original PF. The offline batch optimization further constrains ensemble members to the regions in which the observed runoff ratio and baseflow index can be reasonably reproduced. This additional constraint by the offline batch optimization significantly improves KGE.

Figure 10 demonstrates the difference between HOOPE-PF and PF by showing the timeseries of runoff and $k_q$ (see Table 2). The ensemble size is 100. $S_{state}$ and $S_{para}$ are set to 0.008 and 0.7, respectively. These inflation hyperparameters provide the 50$^{th}$ highest KGE in HOOPE-PF. The purpose of Figure 10 is to demonstrate how different the estimated probabilistic distribution of parameters is in the PF and HOOPE-PF. Note that the comparison of the overall performances of PF and HOOPE-PF is not the purpose of these figures. The superiority of HOOPE-PF over PF has already been shown in Figures 9, S1, S2, and S3. Note that $k_q$ shown in Figures 10c and 10d is a normalized value. Although PF accurately estimates flood peaks in many cases, PF overestimates baseflow (Figure 10b) due to the unstable behavior of the estimated parameters (Figure 10d). This behavior of PF can also be found even when the optimal hyperparameters were chosen to maximize KGE of PF (Figures S1 and S2). On the other hand, HOOPE-PF successfully simulates both high flow and baseflow (Figure 10a). The temporal change of $k_q$ in HOOPE-PF is stabler than that in the original PF (Figures 10c and 10d). In HOOPE-PF, $k_q$ rapidly increases mainly in the flood periods. It is generally difficult for hydrological models with a single set of parameters to accurately simulate runoff in both wet and dry periods (e.g., Deb and Kiem 2020). The application of HOOPE-PF to HYMOD overcomes this limitation of hydrological modeling by allowing model parameters to abruptly change based on the observed runoff. It should be noted that the original PF can also track the abrupt changes in hydrological parameters although it needs the intensive hyperparameter turning and/or the larger ensemble size, as shown in Figure 9. Figure 9 clearly shows that the parameter perturbation method in HOOPE-PF significantly improves the skill to successfully track the abrupt changes in model parameters.

## 5. Conclusions and discussion

In this paper, the new efficient and practical method to estimate model parameters allowing them to temporally change was proposed by combining the offline batch optimization and the online data assimilation. In the newly proposed method, HOOPE-PF, the estimated model parameters by the sequential data assimilation are constrained to the result of the offline batch optimization in which the posterior distribution of model parameters is obtained by comparing the simulated and observed climatological variables. HOOPE-PF outperforms the original sampling-importance-resampling particle filter in the synthetic experiments with the Lorenz 63 model and the real-data experiment of the rainfall-runoff analysis especially when the ensemble size is small, which shows the efficiency of HOOPE-PF. The advantage of HOOPE-PF is that the performance of the online data assimilation is not greatly affected by the "inflation" hyperparameter for model parameters, so that the extensive tuning of this hyperparameter is not necessary. Considering the true dynamics of time-varying parameters is unknown, it is difficult to determine which hyperparameters are accurate in real-world applications, which makes the tuning of hyperparameters complicated. In this case, the stability of HOOPE-PF contributes to improving the efficiency and practicality to estimate convincing time-varying parameters. The efficiency and stability of HOOPE-PF may contribute to solving computationally expensive problems related to time-varying model parameters such as large sample studies on conceptual hydrological models (e.g., Knoben et al. 2020). The inflation of model parameters has been intensively discussed in the previous works on data assimilation (e.g., Moradkhani et al. 2005a, 2005b; Moradkhani et al. 2012; Ruiz et al. 2013; Kotsuki et al. 2019), and HOOPE-PF contributes to these published literatures.

The proposed HOOPE-PF can be the practical approach to detect and quantify unmodelled processes such as land use change (e.g., Pathiraja et al. 2018) from long-term observation. In addition, modeling with time-varying parameters realized stochastic process models whose potential was intensively discussed by Reichert et al. (2021). Reichert et al. (2021) clearly shows that the stochastic process models with time-varying parameters have the advantage to identify an intrinsic model uncertainty and improve model processes.

Since PF is robust to non-linear dynamics and non-Gaussian statistics, it has been widely applied to analyze and predict non-linear earth systems such as land (e.g., Qin et al. 2009; Abolafia-Rosenzweig et al. 2019), ecosystem (e.g., Sawada et al. 2015), paleoclimate reconstruction (e.g., Mairesse et al. 2013), and hydrology (e.g., Moradkhani et al. 2005). The proposed HOOPE-PF can contribute to improving the simulation of these earth systems by modeling time-varying parameters. In addition, the surrogate modeling-based offline batch optimization has been successfully implemented in many earth system models such as atmospheric models (e.g., Dunber et al 2021; Duan et al. 2017; Qian et al. 2018), hydrological models (e.g., Parente et al. 2019; Zhang et al. 2020), and land surface models (e.g., Sawada 2020). Therefore, HOOPE-PF has a potential to be used for a wide variety of earth system models.

There are three limitations in HOOPE-PF. First, since HOOPE-PF requires the long-term observation before beginning the real-time online estimation, it cannot be applied to some types of the real-time estimation problem in which there is no observation at the initial time and the estimation of state variables and parameters needs to be gradually improved whenever new observations are obtained. This situation is often assumed in the previous works about data assimilation on groundwater modelling (e.g., Hendrick-Franssen et al. 2008; Ramgraber et al. 2019) although the long-term observation timeseries can be obtained in many geoscientific applications such as numerical weather prediction and rainfall-runoff analysis. In addition, if long-term observations do not cover the information about "true" parameters for future dynamics, HOOPE-PF cannot accurately estimate them since the future 'true' parameters are located outside the probability distribution estimated by the offline batch optimization. Climate change is one of the possible examples. It should be noted that even in this case HOOPE-PF can evaluate how far the online estimation of parameters are located from the mode of the probability distribution estimated by the offline batch optimization and objectively detect the unprecedented situation which cannot be explained in previous records.

Second, there is no generally applicable way to construct the appropriate "climatological index" for the offline batch optimization (equation (15)). Equation (15) needs to be the summary of the climatology of the important process for a user's purpose. It should be independent to the initial condition. The design of equation (15) apparently depends on the problems. The design of the appropriate climatological index for the offline batch optimization is generally challenging (e.g., Springer et al. 2021). It is promising that the climatological index based only on the widely used hydrological indices, runoff ratio and baseflow index, appropriately works in the application of HOOPE-PF to the conceptual hydrological model. It implies that no complicated methods are needed to construct equation (15).

Third, the current HOOPE-PF cannot be directly applied to high-dimensional problems because the number of required ensemble members increases exponentially with the size of problems (e.g., Snyder et al. 2008, 2015). The recent progress on the application of PF to the high-dimensional problem (i.e. numerical weather prediction) (e.g., Poterjoy et al. 2019; Kawabata and Ueno 2020, see also van Leeuwen et al. 2019 for the comprehensive review) needs to be included to directly apply HOOPE-PF to optimize parameters of high-dimensional models. In addition, the fundamental idea of HOOPE-PF, which is the combination of online data assimilation and offline batch optimization to stabilize the estimation of time-varying parameters, can be transferred to EnKF which has been successfully applied to the high-dimensional problems. The future work will be to develop the version of ensemble Kalman filter (HOOPE-EnKF) based on the similar strategy which is shown in this paper.

**Acknowledgements**

The CAMELS dataset can be downloaded at https://ral.ucar.edu/solutions/products/camels. I thank Ken Dixon for providing a useful Python source code of the Lorenz63 model which was used in this paper and can be downloaded at https://github.com/kendixon/Lorenz63. I also thank Kel Markert for providing a useful Python source code of the HYMOD model which was used in this paper and can be downloaded at https://github.com/KMarkert/hymod. The study was supported by the JST FOREST program (grant no. JPMJFR205Q), the KAKENHI grant (grant no. 21H01430), and the Foundation of River & basin Integrated CommunicationS (FRICS). I thank two anonymous reviewers for their helpful comments.

**Table 1.** Summary and detailed settings of the case studies in this paper

| Name | Model | ensemble size | $S_{state}$ | $S_{para}$ |
|---|---|---|---|---|
| Case Study I | Lorenz63 with a time-varying parameter (equation (23)) | [30,100,250] | [0.15, 0.175, 0.2, 0.225, 0.25, 0.275, 0.3, 0.325, 0.35, 0.375] | [0.1, 0.2, 0.3, 0.4, 0.5, 0.6, 0.7, 0.8, 0.9, 1.0] |
| Case Study II | Lorenz63 with a time-varying parameter (equations (24-25)) | [30,100,250] | [0.15, 0.175, 0.2, 0.225, 0.25, 0.275, 0.3, 0.325, 0.35, 0.375] | [0.1, 0.2, 0.3, 0.4, 0.5, 0.6, 0.7, 0.8, 0.9, 1.0] |
| Case Study III | HYMOD | [30,100,250] | [0.001, 0.002, 0.003, 0.004, 0.005, 0.006, 0.007, 0.008, 0.009, 0.01] | [0.1, 0.2, 0.3, 0.4, 0.5, 0.6, 0.7, 0.8, 0.9, 1.0] |

**Table 2.** Parameters of the HYMOD model.

| Symbols | Description | Units | Range |
|---|---|---|---|
| $C_{max}$ | Maximum soil storage depth | [mm] | [10, 8000] |
| $b_{exp}$ | Pareto-distributed soil storage shape parameter | [-] | [0.1, 2.0] |
| $\alpha$ | Excess runoff splitting parameter | [-] | [0.01, 0.99] |
| $k_s$ | Slow flow routing coefficient | [-] | [0.001, 0.200] |
| $k_q$ | Quick flow routing coefficient | [-] | [0.200, 0.990] |

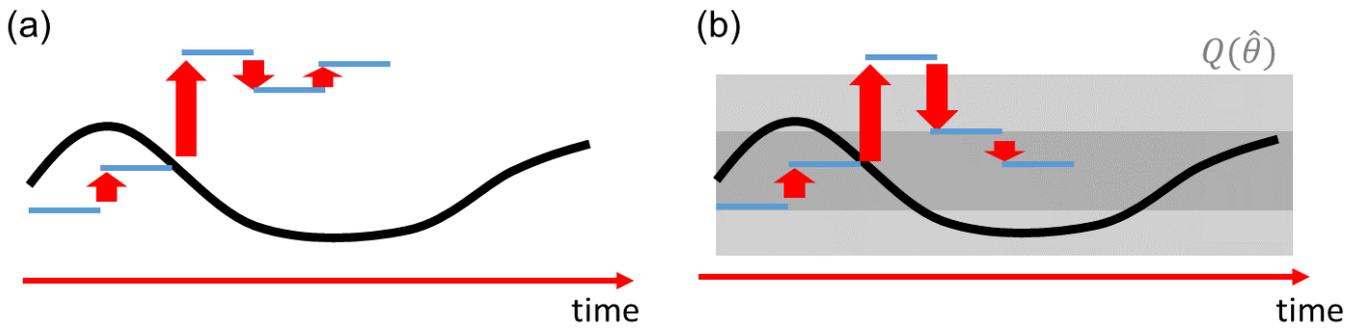

**Figure 1.** Schematics of the (a) original PF and (b) HOOPE-PF. Black lines are timeseries of the true model parameter. Blue lines are the estimation of the online data assimilation algorithm and red arrows show the adjustments by the data assimilation steps. The grey area in (b) shows the posterior distribution of model parameters calculated by the offline batch optimization. When estimated parameters substantially deviate from the truth due to erroneous observations and/or sampling error with insufficient ensemble size, there is a risk of filter degeneracy. HOOPE-PF is expected to prevent filter degeneracy in this situation by efficiently move particles to Q(.) which is the probabilistic distribution of time-invariant parameters. See also section 2.3.

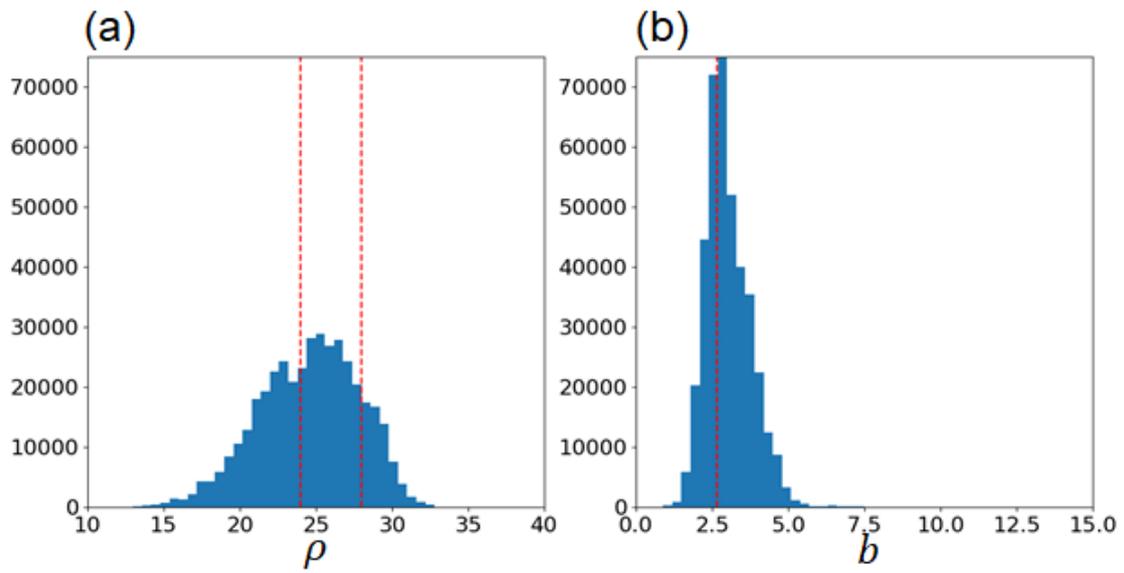

**Figure 2.** Histograms of (a) $\rho(t)$ and (b) $b$ of the Lorenz 63 model estimated by the MCMC sampler in the case study I (see sections 3.1. and 4.1). Red dashed lines are the synthetic truth of the model parameters.

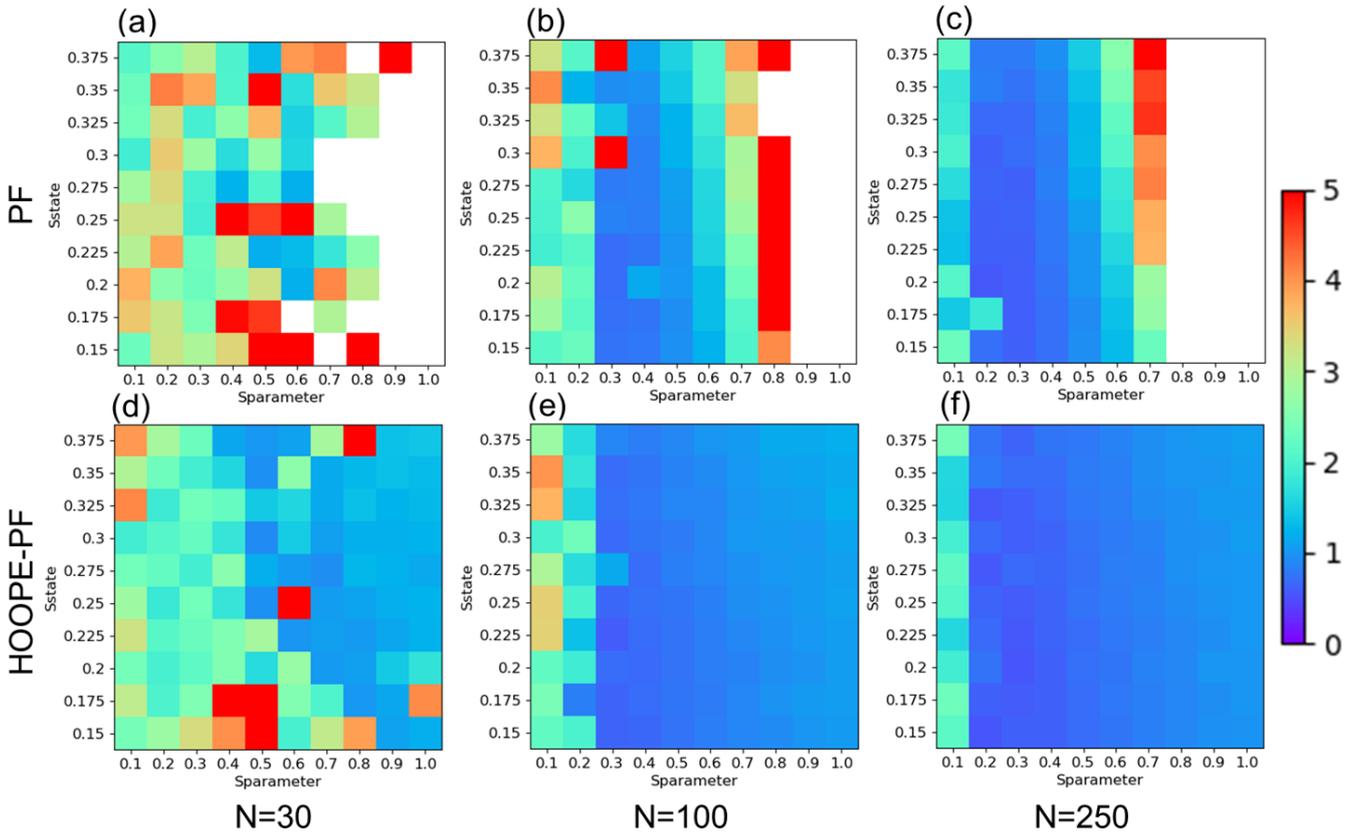

**Figure 3.** (a-c) RMSE between the synthetic true $\rho(t)$ and the ensemble median $\rho(t)$ estimated by the original PF with the ensemble size of (a) 30, (b) 100, and (c) 250 in the case study I (see sections 3.1 and 4.1). Horizontal and vertical axes show $S_{para}$ and $S_{state}$, respectively. (d-f) same as (a-c) but for HOOPE-PF.

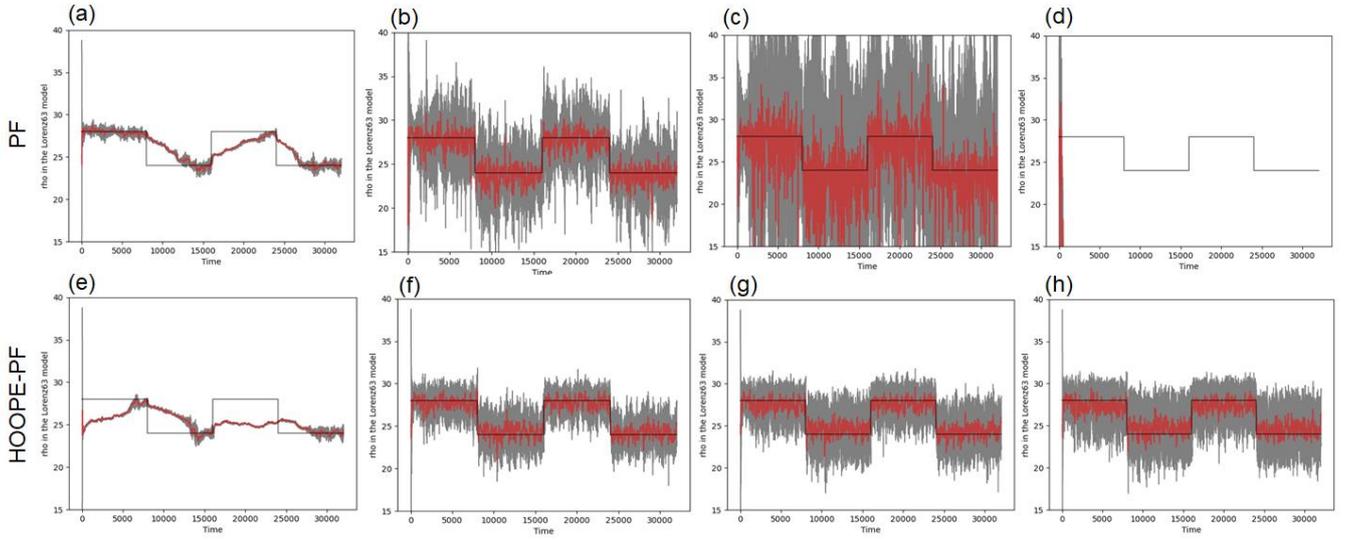

**Figure 4.** (a-d) Timeseries of $\rho(t)$ by the original PF in the case study I (see sections 3.1 and 4.1). Black and red lines show the synthetic truth and the median of the estimated parameter by the online data assimilation, respectively. Grey areas show the 5-95 percentile range. The ensemble size is set to 250. $S_{state}$ is set to 0.25. $S_{para}$ is set to (a) 0.1, (b) 0.5, (c) 0.7, and (d) 0.9. (e-h) same as (a-d) but for HOOPE-PF.

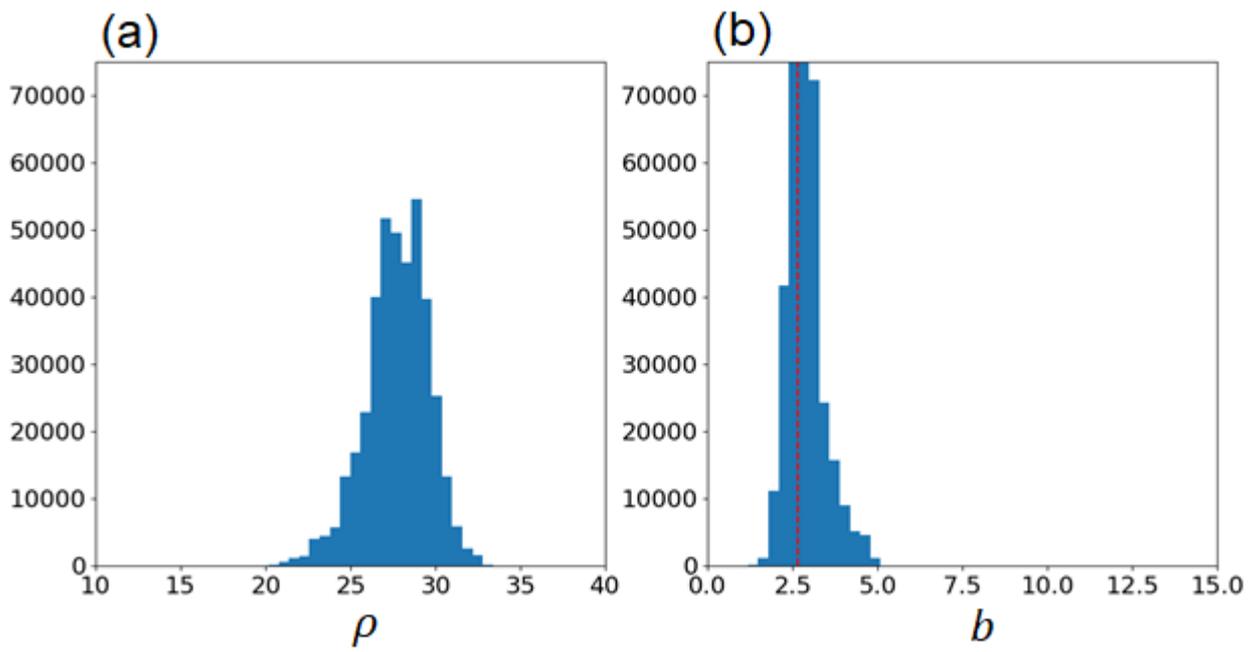

**Figure 5**. Histograms of (a) $\rho(t)$ and (b) $b$ of the Lorenz 63 model estimated by the MCMC sampler in the case study II (see sections 3.2. and 4.2). Red dashed line is the synthetic truth of the model parameters.

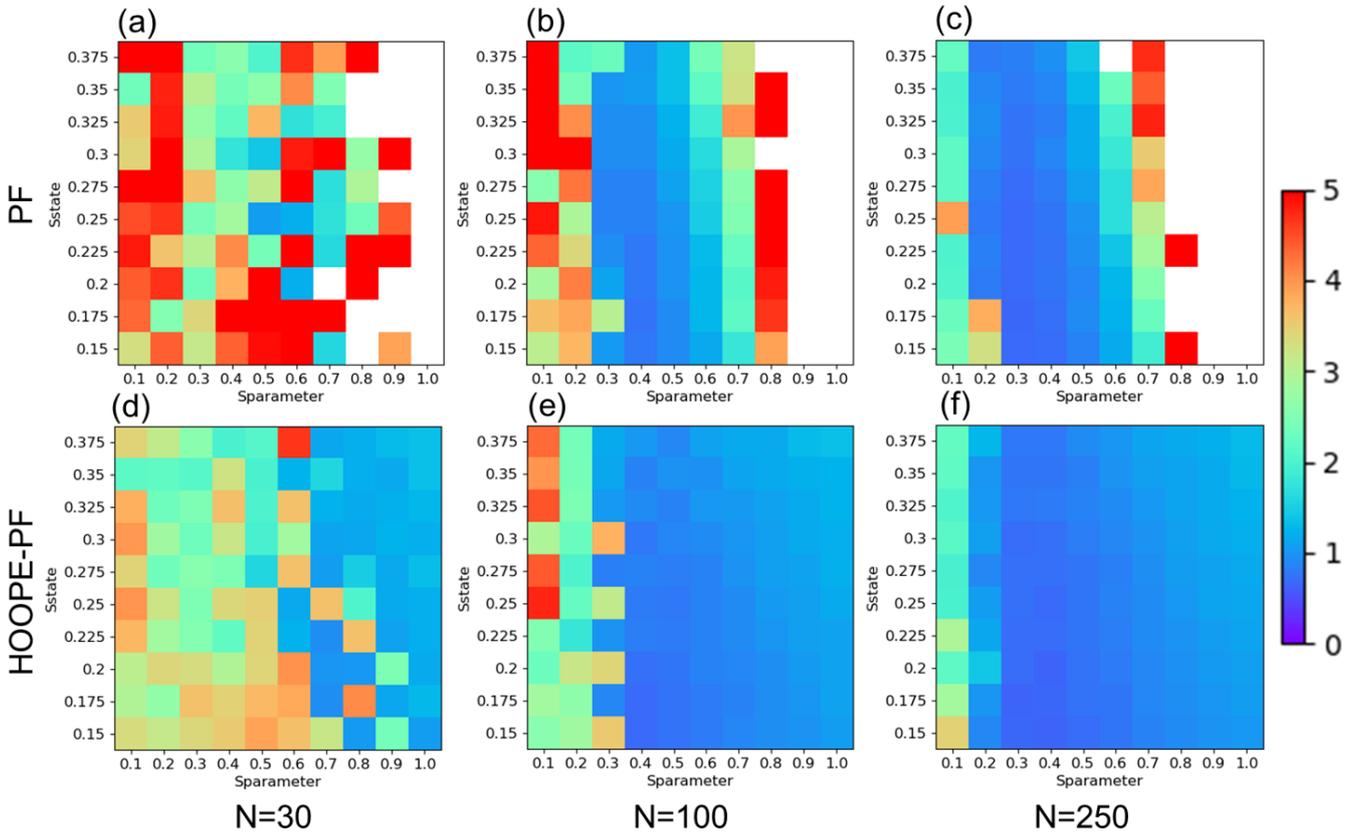

**Figure 6.** (a-c) RMSE between the synthetic true $\rho(t)$ and the ensemble median $\rho(t)$ estimated by the original PF with the ensemble size of (a) 30, (b) 100, and (c) 250 in the case study II (see sections 3.2 and 4.2). Horizontal and vertical axes show $S_{para}$ and $S_{state}$, respectively. (d-f) same as (a-c) but for HOOPE-PF.

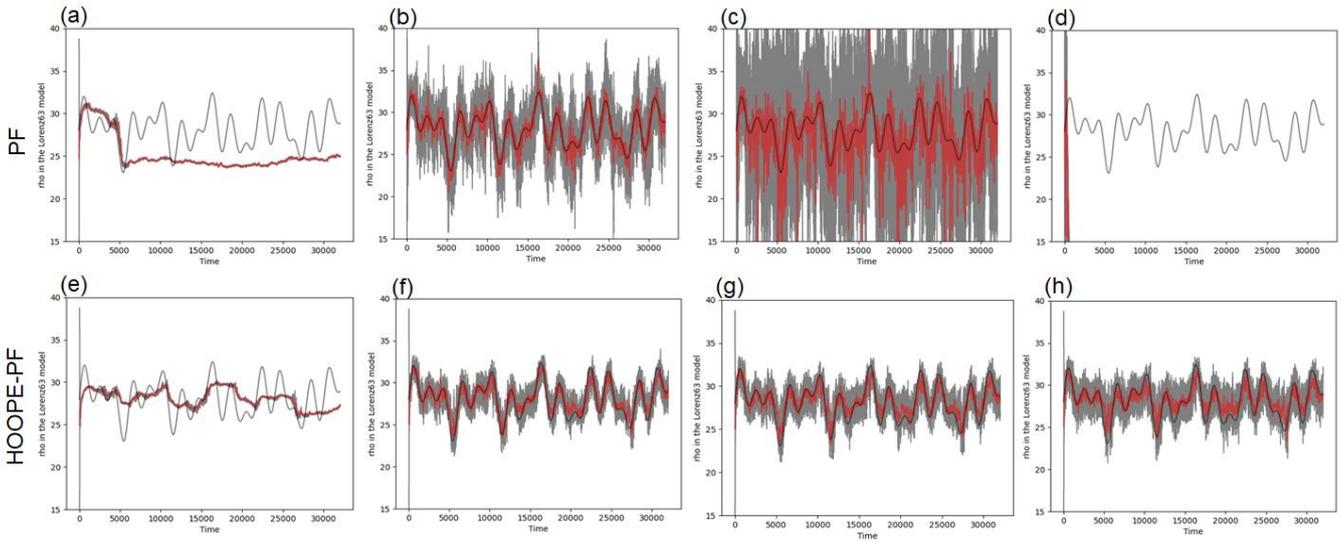

**Figure 7.** (a-d) Timeseries of $\rho(t)$ by the original PF in the case study II (see sections 3.2 and 4.2). Black and red lines show the synthetic truth and the median of the estimated parameter by the online data assimilation, respectively. Grey areas show the 5-95 percentile range. The ensemble size is set to 250. $S_{state}$ is set to 0.25. $S_{para}$ is set to (a) 0.1, (b) 0.5, (c) 0.7, and (d) 0.9. (e-h) same as (a-d) but for HOOPE-PF.

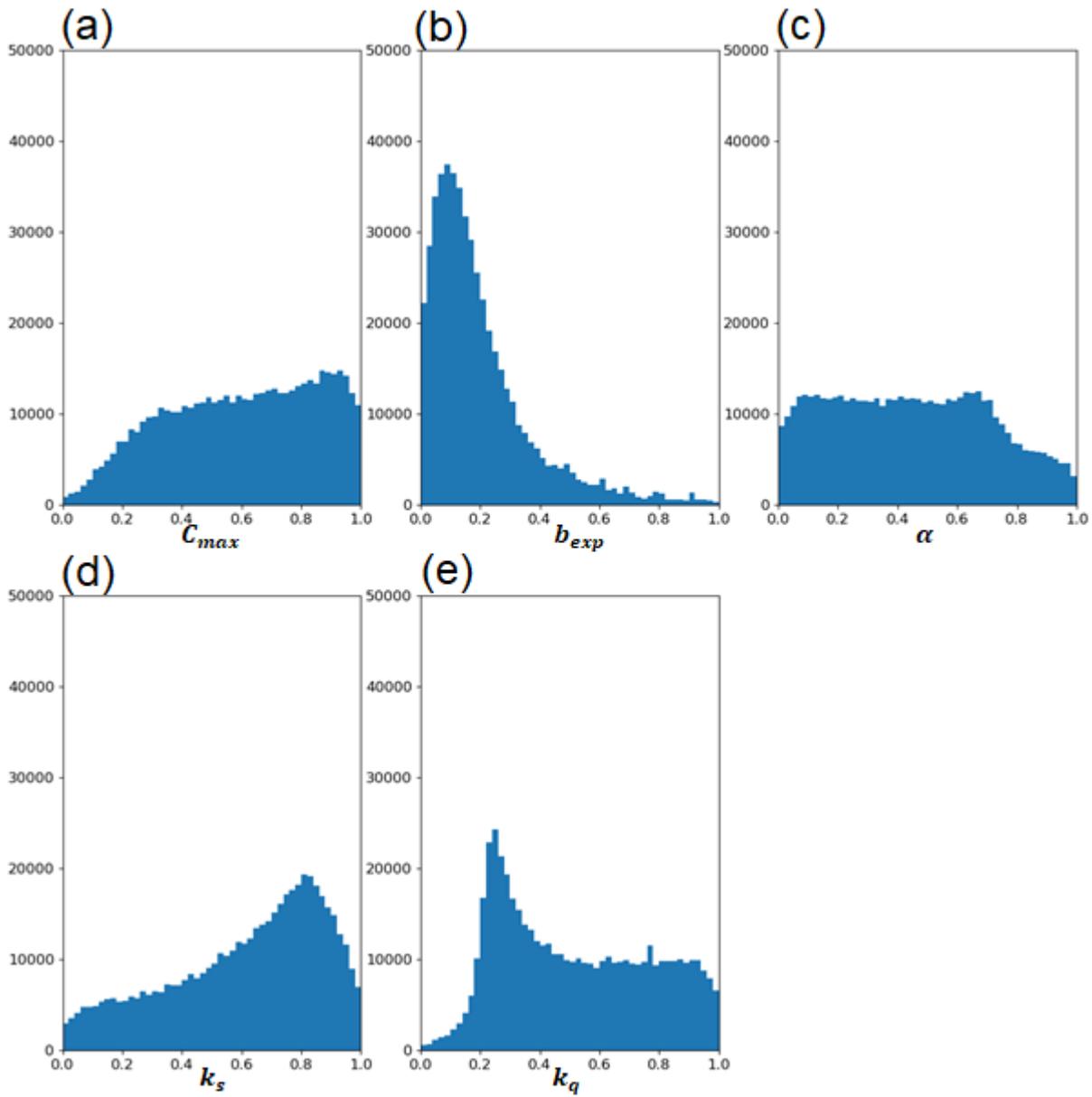

**Figure 8**. Histograms of normalized (a) $C_{max}$, (b) $b_{exp}$, (c) $\alpha$, (d) $k_s$, and (e) $k_q$ of the HYMOD model estimated by the MCMC sampler in the case study III (see sections 3.3. and 4.3).

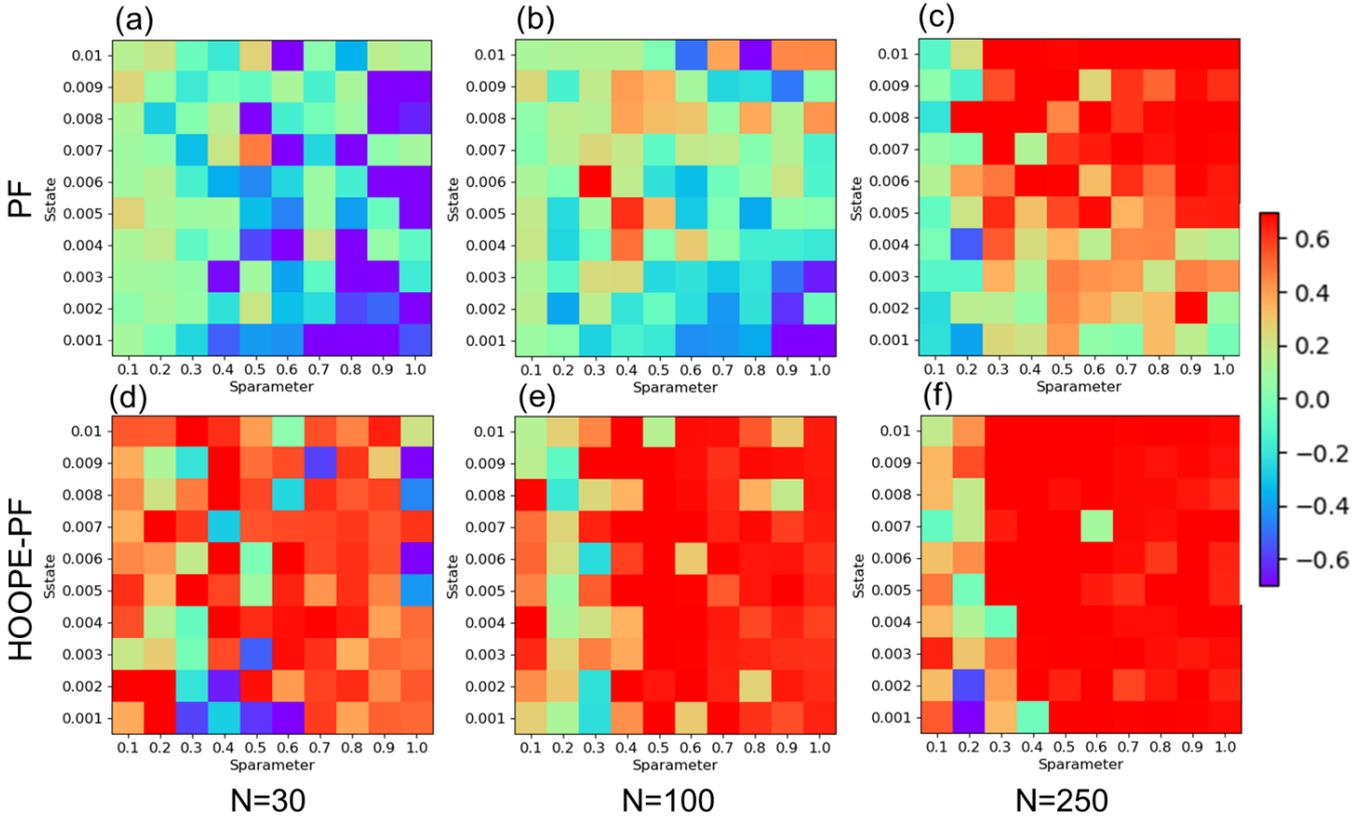

**Figure 9**. (a-c) KGE between the observed and the estimated runoff in the Leaf River by the original PF with the ensemble size of (a) 30, (b) 100, and (c) 250 in the case study III (see sections 3.3 and 4.3). The ensemble median is used as the simulated runoff. Horizontal and vertical axes show $S_{para}$ and $S_{state}$, respectively. (d-f) same as (a-c) but for HOOPE-PF.

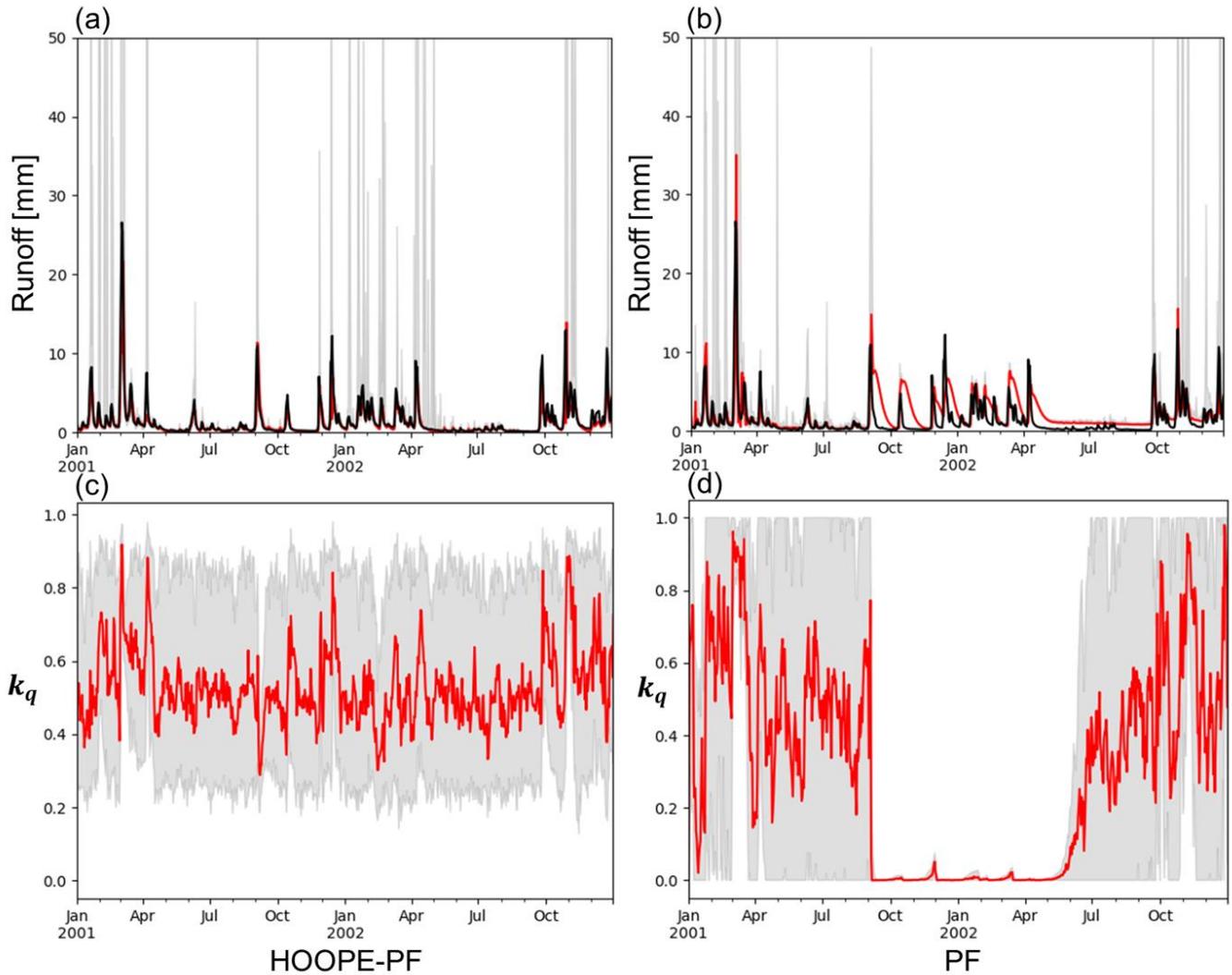

**Figure 10**. (a-b) Timeseries of daily discharge in the Leaf River basin. Black line shows the observed river discharge. Red lines and grey areas show median and the 10-90 percentile range of simulated river discharge by (a) HOOPE-PF and (b) PF. (c-d) Timeseries of the normalized quick flow routing coefficient (see Table 2) in the Leaf River basin. Red lines and grey areas show median and the 10-90 percentile range of the simulated normalized quick flow routing coefficient by (c) HOOPE-PF and (d) PF. The ensemble size is 100. $S_{state}$ and $S_{para}$ are set to 0.008 and 0.7, respectively. These inflation hyperparameters provide the 50$^{th}$ highest KGE in HOOPE-PF.